\documentclass[prX,showpacs,superscriptaddress,twocolumn ]{revtex4}

\usepackage{graphicx}
\usepackage{amsmath}
\usepackage{amsfonts}
\usepackage{amssymb}
\usepackage{epsf}
\usepackage{hyperref}

\begin{document}
\title{Optically guided beam splitter for propagating matter waves}

\author{G. L. Gattobigio}
\affiliation{Laboratoire de Collisions Agr\'egats R\'eactivit\'e,
CNRS UMR 5589, IRSAMC, Universit\'e Paul Sabatier, 118 Route de
Narbonne, 31062 Toulouse CEDEX 4, France}
\affiliation{Laboratoire Kastler Brossel, Ecole Normale
Sup\'erieure, 24 rue Lhomond, 75005 Paris, France}

\author{A. Couvert}
\affiliation{Laboratoire Kastler Brossel, Ecole Normale
Sup\'erieure, 24 rue Lhomond, 75005 Paris, France}

\author{G. Reinaudi}
\affiliation{Laboratoire Kastler Brossel, Ecole Normale
Sup\'erieure, 24 rue Lhomond, 75005 Paris, France}
\affiliation{Department of Physics, Columbia University, 538 West 120th Street, New York, NY 10027}

\author{B. Georgeot}
 \affiliation{Laboratoire de Physique Th\'eorique (IRSAMC), Universit\'e de Toulouse 
(UPS), 31062 Toulouse, France} 
\affiliation{CNRS, LPT UMR5152 (IRSAMC), 31062 Toulouse, France}

\author{D. Gu\'ery-Odelin}
\affiliation{Laboratoire de Collisions Agr\'egats R\'eactivit\'e,
CNRS UMR 5589, IRSAMC, Universit\'e Paul Sabatier, 118 Route de
Narbonne, 31062 Toulouse CEDEX 4, France}
\affiliation{Laboratoire Kastler Brossel, Ecole Normale
Sup\'erieure, 24 rue Lhomond, 75005 Paris, France}

 \date{\today}

\begin{abstract}
We study experimentally and theoretically a beam splitter setup for guided atomic matter waves. The matter wave is a guided atom laser that can be tuned from quasi-monomode to a regime where many transverse modes are populated, and propagates in a horizontal dipole beam until it crosses another horizontal beam at 45$^{\rm o}$. We show that depending on the parameters of this $X$ configuration, the atoms can all end up in one of the two beams (the system behaves as a perfect guide switch), or be split between the four available channels (the system behaves as a beam splitter). The splitting regime results from a chaotic scattering dynamics. The existence of these different regimes turns out to be robust against small variations of the parameters of the system. From numerical studies, we also propose a scheme that provides a robust and controlled beam splitter in two channels only.
\end{abstract}


\maketitle

Matter-wave interferometry has the potential of being several orders of magnitude more sensitive than optical interferometry. 
Many demonstrations have been made of its use as high-precision gravimeters or gyrometers \cite{CSP09}. Its development requires the coherent manipulation of matter waves with atom optical elements. In this respect, the achievement
of beam splitters for atoms moving in free space was essential \cite{BSA84,MOM88,KaC91,Ant06,DDJ08,CGN09,MCH09}.  

To optimize the sensitivity of an atomic interferometer, one needs large angle beam splitters with atom wave packets into superposition of two narrow momentum distribution with large momentum separation. Narrow distribution yields good fringe contrast \cite{WWD05}.
A few strategies are currently explored or envisioned to increase the enclosed area: the enhancement of the interaction time based on large momentum beam splitter \cite{CGN09,MCH09,CKC11}, atom interferometry experiments placed in a reduced gravity environment \cite{LPA98,SDA01}, or the use of slow atoms in a guided environment. 

In this Letter, we shed light on the physics of matter wave splitting in a crossing guide configuration. Confined geometry for atom interferometry are promising in terms of compactness and portability.
The wide variety of techniques available to design potentials and guides for the external degrees of freedom 
has been used to investigate the operation mode of different type of beam splitters. One should distinguish those involving a trap \cite{WAB05,SHA05,HHS05,GDH06,JSW07,MMS10,CKC11} and those 
based on combinations of waveguides \cite{CHF00,MCP00,MCP01,HKP00,DMV02}. So far, the former have been investigated with Bose-Einstein condensate (BEC) while
the latter were studied in the thermal regime. In this work we report on the study of an all-optical beam splitter structure for propagating matter waves both experimentally and numerically and in both regimes (i.e. various initial transverse modes of the guide from the multimode to the monomode limit \cite{GCJ09}). This problem amounts to a quantum scattering problem in confined environment in which chaotic behavior can emerge \cite{AVS09,NJP10,GCG11}.

\begin{figure}[b]
\centerline{\includegraphics[width=6cm]{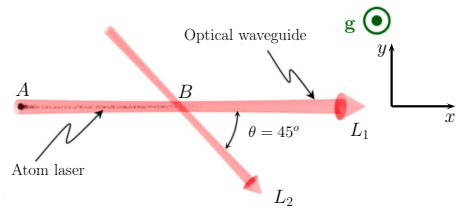}}
 \caption{(color online) Schematic of the matter wave guided beam splitter setup based on two crossing dipole beams in a horizontal plane. Atoms are outcoupled from a BEC located at point $A$ and propagate in the beam $L_1$ towards point $B$ where beam $L_2$ crosses beam $L_1$.} \label{fig1}
\end{figure}

The beam splitter is obtained by combining two guides which provide a potential structure where four paths are available for the matter wave. 
In our experimental setup, we use two dipole beams crossing at 45$^{\rm o}$ in an horizontal plane (see Fig.~\ref{fig1}). This $X$ configuration has four channels \cite{BoB04}.
Optical confinement has many advantages: it can be applied to atoms with no permanent magnetic moment (e.g., ytterbium and alkaline-earth), which is particularly interesting for metrology, an extra magnetic field can be used to tune the interactions (Feshbach resonance) and the focus position can be easily moved \cite{GCL01,CKR08}. Furthermore, the crossing region has a confinement strength larger than that of the guides. As a result the setup is robust against low frequency noise. This is to be contrasted with other Y geometries where the connection zone between the guides implies an important lowering of the potential strength \cite{CHF00,MCP00,MCP01,JaS02}.

We produce a rubidium-87 BEC in a crossed dipole trap (located at $A$ in Fig.~\ref{fig1}) using two focussed laser beams (wavelength of 1070 nm, waist of $w_1\simeq$ 35 $\mu$m) \cite{GCJ09}. Beams $L_1$ and $L_2$ originate from the same laser but have a frequency difference of 80 MHz to wash out possible interferences. Atoms are prepared in the $|F=1,m_F=0\rangle$ internal state through a spin distillation process implemented during the evaporative cooling stage \cite{CJK08}. The guided atom lasers are realized by outcoupling atoms from the BEC in the horizontal dipole beam of the trap either by applying a time-dependent magnetic field gradient or by reducing the intensity of the non horizontal beam of the crossing dipole trap \cite{GRG06,CJK08,GCJ09,KWE10, DHJ10, DHM11}. 

Although atom-atom interactions may not always alter dramatically the coherence \cite{DRL10}, theoretical studies of cold atoms interferometers based on guiding potentials show that they can affect dramatically the contrast of the expected interference signal \cite{StZ02,ChE03}. To reduce the role played by interactions we perform the outcoupling very slowly. The low atomic density of the propagating beam makes this system nearly interaction-free \cite{CJK08,GCJ09}. The propagating matter wave is charaterized by its atom flux ($\sim$ few 10$^5$ atoms/s), its mean velocity ($13\pm 2$ mm.s$^{-1}$) and the mean excitation number $\langle n \rangle$ associated to the transverse modes.

Experiments are carried out at a fixed power $P_1=140$ mW of the guide beam $L_1$ which corresponds to a measured transverse frequency of about 300 Hz and a depth of 10 $\mu$K. The crossing beam $L_2$ has a waist $w_2=$ 70$\pm$5 $\mu$m and an adjustable power $P_2$. 
Atoms propagate for a time long enough to obtain clear information on the asymptotic output channels. In practice, the crossing between $L_1$ and $L_2$ beams is at 700 $\mu$m downstream from the BEC. The beam propagates during 200 ms, and we take an absorption picture after a few ms time-of-flight. 

\begin{figure}[b]
\centerline{\includegraphics[width=\columnwidth]{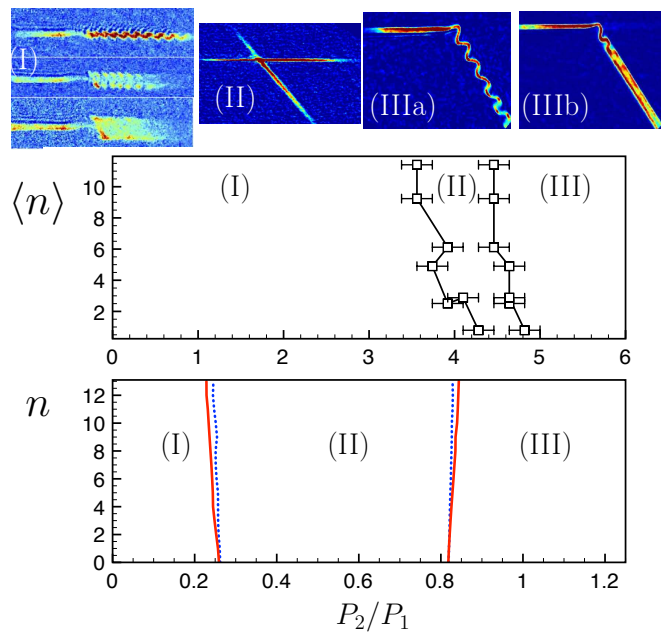}}
 \caption{(color online) Propagation of an atom laser in a X shape guide structure. Top: Three regimes are experimentally observed depending on the power ratio associated with each guide: (I) weak defect regime where the excitation of transverse modes downstream from the crossing increases with the power ratio $P_2/P_1$, (II) splitting regime where atoms explored all the available channels, and (IIIa/b)  switch regime where all atoms are directed into guide $L_2$. The mean transverse excitation numbers of the incoming beam are $\langle n \rangle \simeq 0.3$ (IIIa) and $\langle n \rangle \simeq 10$ (IIIb). Middle: Experimental diagram that summarizes the observed regime depending on both the power ratio $P_2/P_1$ and the mean transverse populations. The boundaries of the three regimes turn out to be robust against the population of the transverse modes. Bottom: Numerical simulations confirming this robustness; red solid line: quantum simulations, blue dotted line: classical simulations. Wave packets are started 150 $\mu$m before the crossing,
with initial longitudinal velocity 10 mm.s$^{-1}$,
for $w_1=45$ $\mu$m, $w_2=41$ $\mu$m.} \label{fig2}
\end{figure}

In a first set of experiments, we prepare the propagating guided atom laser in the ground state of the transverse confinement ($\langle n \rangle \simeq 0$). In this transverse monomode regime, we observe three different regimes (I), (II) and (III) depending on $P_2/P_1$ (see Fig.~\ref{fig2}). Images (I) belong to the weak defect regime for which the beam is transversally excited after passing through the interaction region. The coherent excitation of the transverse modes is induced by the weak coupling between the longitudinal and transverse degrees of freedom in the crossing guide region. As illlustrated in Fig.~\ref{fig2}, the larger $P_2/P_1$, the stronger the coupling and the larger the number of transverse modes populated downstream from the crossing region. Image (II) corresponds to the splitting regime where atoms can explore the four channels provided by our X shape geometry. It  corresponds to a much stronger coupling between the different degrees of freedom and the corresponding classical dynamics becomes chaotic (see below) \cite{TER10,GCG11}. Images (III) illustrate the switch regime where atoms go into one arm of the beam $L_2$ with a 100 \% efficiency \cite{MCP00, MCP01}. The center of mass motion is excited as a result of momentum transfer of the longitudinal incident momentum on the transverse degrees of freedom of the output channel guide. 
 We plot a boundary between (I) and (II)  in Fig.~\ref{fig2} (middle) when less than 80 \% of the atoms continue their propagation in the original guide $L_1$, and between (II) and (III) when more than 80 \% of the atoms are deviated to the guide $L_2$. The error bars show the typical size of the transition regions.

 We also have investigated experimentally this dynamics for various values of the initial mean quantum number $\langle n \rangle$ of the transverse states by adjusting the fraction of condensed atoms from which the guided atom laser is produced \cite{GCJ09}. For $\langle n \rangle=10$ more than hundreds of transverse levels are populated since one has to take into account the degeneracy, while one transverse state is populated for $\langle n \rangle=0$. The results are summarized in Fig.~\ref{fig2} (middle). We observe that the three regimes are still present. However, the transverse oscillations can be observed only for low incoming mean transverse excitation (compare images (IIIa) and (IIIb) taken for different values of $\langle n \rangle$). In previous experiments where the splitting of a beam of cold atoms was carried out, it was impossible to observe this transverse excitation since many transverse states were populated and these oscillations were thus washed out \cite{CHF00,MCP00,HKP00,DMV02}. The boundaries between the different regimes appear to be very robust against $\langle n \rangle$ and a slight misalignement of the crossing beam. This is to be contrasted with the interaction of a guided atom laser with an out-of-center vertical potential \cite{GCG11}. The boundaries remain nearly unchanged when the crossing is misaligned by a tenth of the waist of the beams. For each set of data the alignment was checked before and after the data acquisition. We estimate that the maximum relative change of the beam positions is below 2 $\mu$m. 

To interpret the splitting regime, we have computed the dynamics of 2D quantum wavepackets using the split-Fourier algorithm \cite{note} and developed a direct simulation Monte Carlo (DSMC) method where atoms are evolved according to classical mechanics (see Fig.~\ref{fig2} (bottom)). The three regimes observed experimentally can be reproduced by numerical simulations and turn out to be generic. Each point in Fig.~\ref{fig2} (bottom) corresponds to the results obtained for an initial state that coincides with the $n^{\rm th}$ transverse eigenstate. We observe that the boundaries deduced from classical and quantum simulations are very close. The fact that the boundaries are not exactly the same in classical and quantum dynamics is somehow expected since the equations fulfilled by the center of mass of the packet are different in classical and quantum mechanics. Indeed, the external potential is not anymore harmonic in the crossing guide region. The error bars on the experimental data are such that one cannot distinguish the classical or  quantum nature of the motion at low $n$. 
Numerics confirm the robustness of the existence of the boundaries. However, their precise positions can depend on several parameters such as small misalignment of the crossing beams, a residual longitudinal acceleration and/or the relative waist size of the two beams. Many of these parameters cannot be measured accurately. Consequently and to avoid the use of too many free parameters, we do not fit the experimental data with the numerics but use the numerics to identify the underlying physics and confirm its generic character and its robustness.

\begin{figure}[b]
\centerline{\includegraphics[width=\columnwidth]{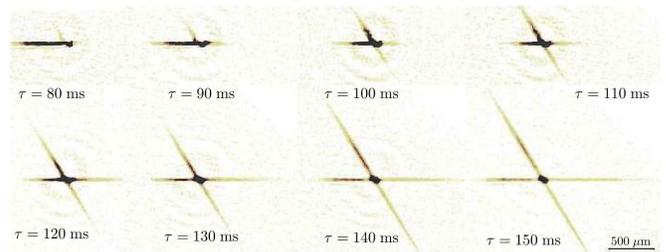}}
 \caption{(color online) Example of experimental splitting dynamics illustrating regime (II) (see Fig.~\ref{fig2}). Time evolution is directly visible in the succession of absorption images.The images show a larger density at the crossing point which is attributed to the complex dynamics that traps the atoms there for some time. Atoms that escape from this region populate all available channels.} \label{fig3}
\end{figure}

\begin{figure}[b]
\centerline{\includegraphics[width=\columnwidth]{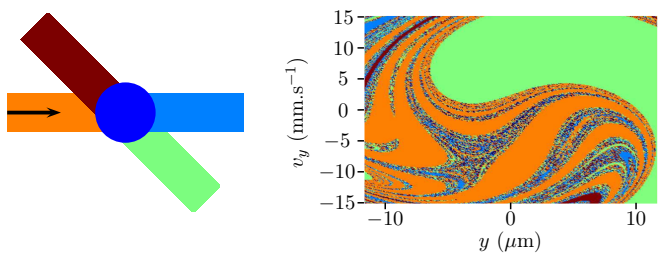}}
 \caption{(color online). Left: colors associated with the different output channels. The arrow indicates the propagation direction of the incoming atoms. The overlap region between the guide is represented as a dark blue disk. Right: Output channels associated to different initial phase space conditions of the transverse degree of freedom in regime (II) (see Fig.~\ref{fig2}). Each classical trajectory started
at the bottom of the guide, 150 $\mu$m before the crossing,
with initial velocity 10 mm.s$^{-1}$,
for $w_1=45$ $\mu$m, $w_2=41$ $\mu$m, $P_2=0.143$ W$=0.8\; P_1$.} \label{fig4}
\end{figure}

Figure \ref{fig3} provides a typical experimental example of the dynamics of the splitting regime. The guided atom laser that enters the crossing region remains partially trapped (larger density) and then the contamination of all available branches is observed. We have plotted in Fig.~\ref{fig4} the output channels obtained from a simulation relying on classical mechanics (using different color for the different output channels) for different initial conditions of the transverse phase space. We clearly identify a chaotic region with a fractal structure in which a very small change in the initial condition changes the output channel. This is the signature of a chaotic dynamics \cite{AVS09}. We have also checked that Lyapunov exponents become positive in this parameter region. To further characterize this zone, we have verified that the set of boundaries between the four accessible output channels as a function of the initial conditions displays a fractal structure. For a wide variety of fixed transverse velocity $v_y$ in regime (II), we find a fractal dimension of $\xi \simeq 0.85$ revealing a strong fractality of the basin boundaries, usually associated with chaotic scattering. Furthermore, numerics seem to indicate that the fractal Wada property \cite{KeY91,SOY99,AVS09} (every point on the boundary of a basin is on the boundary of all basins) can be fulfilled in this system. The distribution of trapping times in the vicinity of the crossing guide region is found to exhibit an exponential decay. 

Interestingly the population in a given channel can be reduced by adding a small extra gradient ; this happens in the experiment if the beams are not perfectly horizontal. In our nearly horizontal configuration, a gradient that corresponds to a hundredth of the gravity strength is sufficient to dissymmetrize dramatically the population in the output channels. In this way one can populate only two channels and realize a two-channel beam splitter. This effect has been checked numerically and explains the variation of the populations in the different channels seen in the experimental results (compare image (II) of Fig.~\ref{fig2} and images of Fig.~\ref{fig3}). A beam splitter experiment for a thermal cloud accelerated by gravity in a quasi-vertical configuration has been discussed in \cite{HKP00,GSP06}.  Note that the possibility of realizing a coherent beam splitter using the transient dynamics of a chaotic system has been addressed in another context with amplitude modulated optical lattices \cite{TFH00}.

The $X$ configuration that we have investigated thus exhibits two regular regions separated by a chaotic region. Actually, the numerical simulations reveal the existence of other splitting region at even higher power ratio $P_2/P_1$ that were not investigated in our experiment. These splitting regimes are undesirable for interferometry since the relative phases of the matter wave in the different output channels can be affected in an uncontrollable manner from run to run. A possible solution as investigated numerically in \cite{KPL04} consists in adding a dedicated extra potential to decrease the coupling between the longitudinal and transverse degrees of freedom.

\begin{figure}[h]
\centerline{\includegraphics[width=8cm]{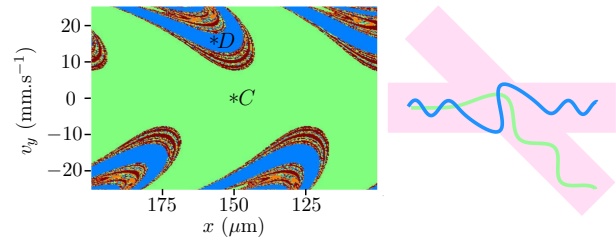}}
 \caption{(color online). Left: Output channels for a classical trajectory started
at the bottom of the guide, between 100 $\mu$m and 200 $\mu$m
before the crossing,
with initial longitudinal velocity 10 mm.s$^{-1}$, and an initial transverse velocity
between $-25$ mm.s$^{-1}$ and 25 mm.s$^{-1}$, with
$w_1=45$ $\mu$m, $w_2=41$ $\mu$m, $P_2=0.365$ W$=2.04\; P_1$.
Right: center of mass motion of numerically simulated quantum wave packets. The one with initial parameters of point $C$ (150 $\mu$m before defect, no transverse velocity) is deviated in the second guide at 99.9 \% (green curve). In contrast, the one with initial parameters of point $D$ (154.75 $\mu$m before defect,
transverse velocity of 15.5 mm.s$^{-1}$) ends up at 96.2 \% in the initial beam (blue curve).} \label{fig5}
\end{figure}

In the following, we propose an alternative scheme using a transverse momentum kick transferred to atoms before they enter the crossing region.
The power ratio is chosen to be in regime (III) (the switch guide regime).
Different experimental techniques can be used to achieve such a transfer: Bragg or Raman transitions, Bloch oscillations \cite{CGO11}. In our simulation, we add an instantaneous transverse momentum upstream to account for the momentum kick. Figure~\ref{fig5} summarizes our numerical results. Each asymptotic output channel obtained here from a classical simulation is represented by a different color as a function of the velocity kick value and the ``phase" i.e. the position before the crossing at which the momentum transfer is performed. Using the prediction of the simulations relying on classical physics as guidelines for quantum simulations, we conclude on the feasibility of re-directing atoms in guide $L_1$ with an efficiency larger than 96 \%. This technique enables to switch from guide $L_2$ to $L_1$ without changing in time the guide characteristics \cite{MCP00,MCP01}.
This also means that if the incoming wave packet is prepared in a linear superposition with equal weight of the states with and without the proper transverse momentum kick, the wave function can be split into two coherent wave packets that propagate in the two guides $L_1$ and $L_2$. Interestingly enough, the domain of useful parameters in momentum kick and phase to realize such a guided matter wave beam splitter is relatively large. This shall ensure the robustness of the method. The splitted beams are for each guide in an excited state that is responsible for their transverse oscillation (see image (III) of Fig.~\ref{fig2}). An out-of-guide axis potential with an appropriate strength can be used to remove the remaining transverse dipole oscillation \cite{GCG11}.

In conclusion, this Letter reports on the realization of all-optical guided beam splitter devices. Such tools are important for the development of guided atomic optics which has a huge potential for applications, in particular for inertial sensors. In addition, it was shown recently that there exists a direct mapping between waveguide theory and spin chain transport \cite{MCH11}, opening possible applications of our scheme in solid state physics.

We thank R. Mathevet and T. Lahaye for useful comments at the early stage of this work.
We acknowledge financial support from the Agence Nationale de la Recherche (GALOP project), the R\'egion Midi-Pyr\'en\'ees, the university Paul Sabatier (OMASYC project), the NEXT project ENCOQUAM and the Institut Universitaire de France. We thank CalMiP for the use of their supercomputers.


\begin{thebibliography}{0}
\expandafter\ifx\csname natexlab\endcsname\relax\def\natexlab#1{#1}\fi
\expandafter\ifx\csname bibnamefont\endcsname\relax
  \def\bibnamefont#1{#1}\fi
\expandafter\ifx\csname bibfnamefont\endcsname\relax
  \def\bibfnamefont#1{#1}\fi
\expandafter\ifx\csname citenamefont\endcsname\relax
  \def\citenamefont#1{#1}\fi
\expandafter\ifx\csname url\endcsname\relax
  \def\url#1{\texttt{#1}}\fi
\expandafter\ifx\csname urlprefix\endcsname\relax\def\urlprefix{URL }\fi
\providecommand{\bibinfo}[2]{#2}
\providecommand{\eprint}[2][]{\url{#2}}

\end{thebibliography}


\begin{thebibliography}{99}


\bibitem{CSP09} A. D. Cronin, J. Schmiedmayer and D. E. Pritchard, Rev. Mod. Phys. \textbf{81}, 1051 (2009).

\bibitem{BSA84}
Ch. J. Bord\'e \emph{et al.} Phys. Rev. A \textbf{30}, 1836 (1984). 

\bibitem{MOM88}
P. J. Martin \emph{et al.}, Phys. Rev. Lett. \textbf{60}, 515 (1988). 

\bibitem{KaC91}
M. Kasevich and S. Chu, Phys. Rev. Lett. \textbf{67}, 181 (1991). 

\bibitem{Ant06}
C. Antoine, Appl. Phys. B \textbf{84}, 585 (2006).

\bibitem{DDJ08}
J. Dugu\'{e} \emph{et al.}, Phys. Rev. A \textbf{77}, 031603 (2008).

\bibitem{CGN09}
P. Clad\'e \emph{et al.} Phys. Rev. Lett. \textbf{102}, 240402 (2009). 

\bibitem{MCH09}
H. M\"uller \emph{et al.}, Phys. Rev. Lett. \textbf{102}, 240403 (2009). 

\bibitem{WWD05}
S. Wu \emph{et al.}, Phys. Rev. A \textbf{71}, 043602 (2005).

\bibitem{CKC11}
S.-w. Chiow \emph{et al.}, Phys. Rev. Lett. \textbf{107}, 130403 (2011).

\bibitem{LPA98}
P. Lemonde \emph{et al.}, Eur. Phys. J. D \textbf{3}, 201 (1998).

\bibitem{SDA01}
C. Salomon \emph{et al.}, C. R. Acad. Sci. \textbf{2}, 1313 (2001).

\bibitem{WAB05}
Y.-J. Wang \emph{et al.} Phys. Rev. Lett. \textbf{94}, 090405 (2005).

\bibitem{SHA05}
T. Schumm \emph{et al.}, Nature Phys. \textbf{1}, 27(2005).

\bibitem{HHS05}
P. Hommelhoff \emph{et al.}, New J. Phys. \textbf{7}, 3 (2005).

\bibitem{GDH06}
O. Garcia \emph{et al.}, Phys. Rev. A \textbf{74}, 031601(R) (2006). 

\bibitem{JSW07}
G.-B. Jo \emph{et al.}, Phys. Rev. Lett. \textbf{98}, 030407 (2007).

\bibitem{MMS10}
K. Maussang \emph{et al.}, Phys. Rev. Lett. \textbf{105}, 080403 (2010).

\bibitem{CHF00}
D. Cassettari \emph{et al.}, Phys. Rev. Lett. \textbf{85}, 5483 (2000).

\bibitem{MCP00}
D. M\"uller \emph{et al.}, Opt. Lett. \textbf{25}, 1382 (2000).

\bibitem{MCP01}
D. M\"uller \emph{et al.}, Phys. Rev. A \textbf{63}, 041602 (2001).

\bibitem{HKP00}
O. Houde, D. Kadio and L. Pruvost, Phys. Rev. Lett. \textbf{85}, 5543 (2000).

\bibitem{DMV02}
R. Dumke \emph{et al.}, Phys. Rev. Lett. \textbf{89}, 220402 (2002).

\bibitem{GCJ09}
G. L. Gattobigio \emph{et al.}, Phys. Rev. A \textbf{80}, 041605(R) (2009).

\bibitem{AVS09}
J. Aguirre, R. L. Viana, and M. A. F. Sanju\'{a}n, Rev. Mod. Phys. \textbf{81}, 333 (2009). 

\bibitem{NJP10} 
G. L. Gattobigio \emph{et al.}, New J. Phys. \textbf{12}, 085013 (2010).

\bibitem{GCG11}
G. L. Gattobigio \emph{et al.}, Phys. Rev. Lett. \textbf{107}, 254104 (2011).

\bibitem{BoB04}
D. C. E. Bortolotti and J. L. Bohn, Phys. Rev. A \textbf{69}, 033607 (2004).

\bibitem{GCL01}
T. L. Gustavson  \emph{et al.}, Phys. Rev. Lett. \textbf{88}, 020401 (2001).

\bibitem{CKR08}
A. Couvert \emph{et al.}, Europhys. Lett. \textbf{83}, 13001 (2008).

\bibitem{JaS02}
M. J\"a\"askel\"ainen and S. Stenholm, Phys. Rev. A \textbf{66}, 023608 (2002) ; \emph{ibid} \textbf{68}, 033607 (2003).

\bibitem{CJK08}
A. Couvert \emph{et al.}, Europhys. Lett. \textbf{83}, 50001 (2008).

\bibitem{GRG06} 
W. Guerin \emph{et al.}, Phys. Rev. Lett. \textbf{97}, 200402 (2006).

\bibitem{KWE10} 
G. Kleine B\"uning \emph{et al.}, Appl. Phys. B \textbf{100}, 117 (2010).

\bibitem{DHJ10}
R. G. Dall \emph{et al.}, Phys. Rev. A \textbf{81}, 011602 (2010). 

\bibitem{DHM11}
R. G. Dall \emph{et al.}, Opt. Lett. \textbf{36}, 1131 (2011).

\bibitem{DRL10}
C. Deutsch \emph{et al.} Phys. Rev. Lett. \textbf{105}, 020401 (2010).

\bibitem{StZ02}
J. A. Stickney and A. A. Zozulya, Phys. Rev. A \textbf{66}, 053601 (2002).

\bibitem{ChE03}
S. Chen and R. Egger, Phys. Rev. A \textbf{68}, 063605 (2003).

\bibitem{TER10}
E. Torrontegui \emph{et al.}, Phys. Rev. A \textbf{82}, 043420 (2010).
 
\bibitem{note}
The large anisotropy of the system (longitudinal size of a few mm and transverse size of a few hundreds of nm) and the long interaction time makes this kind of calculation unrealistic in 3D with the currently available supercomputers.

\bibitem{KeY91}
J. Kennedy and J. A. Yorke, Physica D \textbf{51}, 213 (1991).

\bibitem{SOY99}
D. Sweet, E. Ott, J. A. Yorke, Nature \textbf{399}, 315 (1999).

\bibitem{GSP06}
N. Gaaloul \emph{et al.}, Phys. Rev. A \textbf{74}, 023620 (2006).

\bibitem{TFH00}
A. G. Truscott \emph{et al.}, Phys. Rev. Lett. \textbf{84}, 4023 (2000).

\bibitem{KPL04}
H. Kreutzmann \emph{et al.}, Phys.  Rev. Lett. \textbf{92}, 163201 (2004).

\bibitem{CGO11}
C. Cohen-Tannouji and D. Gu\'{e}ry-Odelin, \textit{Advances in atomic physics: an overview} (World Scientific, Singapore, 2011).

\bibitem{MCH11}
M. I. Makin \emph{et al.}, Phys. Rev. Lett. \textbf{108}, 017207 (2012). 

\end{thebibliography}
\end{document}